\documentclass[cits]{PoS}

\title{GeV Gamma Rays from Supernova Remnants Interacting with Molecular Clouds}

\ShortTitle{GeV Gamma-Rays from SNRs}

\author{\speaker{Yasunobu Uchiyama}%
        \thanks{Panofsky Fellow.}\\
       SLAC National Accelerator Laboratory, Stanford University\\
       E-mail: \email{uchiyama@slac.stanford.edu}}

\author{On behalf of the Fermi-LAT Collaboration}

\abstract{
We report the current status of 
the observations of supernova remnants (SNRs) 
with the Large Area Telescope (LAT) aboard 
the \emph{Fermi} Gamma-ray Space Telescope,  
focusing on 
middle-aged SNRs that appear to be interacting with molecular clouds. 
Observations with the \emph{Fermi} LAT  
in an energy range from 0.2 GeV to $\sim 100$ GeV
have unveiled the presence of luminous GeV gamma-ray emission in middle-aged 
SNRs, providing a new insight into the shock-acceleration theory and the origin of galactic cosmic rays. 
The middle-aged SNRs detected by the \emph{Fermi} LAT 
 are generally much brighter in GeV than in TeV in terms of energy flux, 
 which emphasizes the importance of the GeV observations. 
Spectral steepening in the Fermi-LAT band is commonly found 
for the GeV-luminous SNRs. 
Remarkably, most (if not all) of the GeV-luminous SNRs are  known to be 
 interacting with molecular clouds, 
and they are also  the strong sources of synchrotron radio emission. 
We discuss possible scenarios to explain the enhanced GeV 
gamma-ray emission in the cloud-interacting SNRs. 
Particular emphasis is placed on a scenario 
 in which shock-accelerated cosmic-rays are adiabatically 
compressed and energized as result of radiative cooling behind 
the cloud shock. }

\FullConference{25th Texas Symposium on Relativistic Astrophysics \\
		 December 6-10, 2010\\
		 Heidelberg, Germany}

\begin{document}

\section{Introduction}

Diffuse shock acceleration \cite{BE87} operating at a collisionless shock
 produces nonthermal particles  in a supernova remnant (SNR). 
Remarkable  progress in the observations of shock-accelerated particles 
was made in the last decade, 
especially in the X-ray and TeV $\gamma$-ray bands, 
which has strengthened the  conjecture that shock acceleration in 
SNRs is responsible for the bulk of the galactic cosmic rays (GCRs). 
Synchrotron X-ray emission produced by multi-TeV electrons has been 
detected from most of young historical SNRs \cite{Reynolds08}.
The X-ray measurements by the \emph{Chandra} 
X-ray Observatory with its arcsecond resolution imaging
have shown that particle acceleration to TeV energies 
at strong shocks in young SNRs accompanies significant amplification of 
interstellar magnetic fields \cite{Uchiyama07}.
Direct evidence for the production of multi-TeV particles in shell-type 
SNRs came from 
the TeV $\gamma$-ray observations made with ground-based 
Cherenkov telescopes such as H.E.S.S. \cite{HESS04}, though 
the emission mechanism (``hadronic vs leptonic") remains unsettled. 
Gamma-ray observations at the GeV band have been expected to provide 
complementary information on particle acceleration at the expanding 
shock waves of galactic SNRs.

In the GeV $\gamma$-ray band, possible associations of $\gamma$-ray 
sources with  five radio-bright shell-type SNRs were reported by 
Ref.~\cite{Esposito} 
based on the data taken in 1991--1994 by 
the EGRET instrument onboard the \emph{Compton} 
Gamma Ray Observatory (see also Ref.\ \cite{Torres}).
The two strongest $\gamma$-ray sources that are 
possibly associated with SNRs are the EGRET unidentified sources 
in the directions of 
$\gamma$~Cygni and IC~443, 
both of which are known to be interacting with adjacent molecular clouds. 
If the $\gamma$-ray excesses are due to cosmic-ray interactions with 
a dense cloud, the dominant emission processes are either 
bremsstrahlung by relativistic electrons or $\pi^0$-decay $\gamma$-rays 
resulting from proton-proton hadronic collisions. 
However, the EGRET point-spread function is much wider than 
the spatial extent of most of the SNRs with possible $\gamma$-ray 
associations, which made it difficult to preclude the possibility that 
the $\gamma$-ray sources are due to 
pulsars\footnote{One of the two 
most significant EGRET sources 
(a source in the $\gamma$ Cygni SNR) 
is indeed found to be a $\gamma$-ray pulsar with the \emph{Fermi} LAT 
\cite{Fermi-Pulsars}.}.

The Large Area Telescope (LAT) onboard the \emph{Fermi} Gamma-ray Space 
Telescope \cite{LAT}, which was launched in June 2008, 
has started to explore the GeV $\gamma$-ray sky with much improved 
resolving power and 
unprecedented sensitivity. The observations with the \emph{Fermi}-LAT 
provide 
new methods of addressing unsolved problems in shock acceleration theory 
and the origin of the GCRs. 
Using the LAT data from the first $\sim 1$ yr of observations,
the \emph{Fermi} LAT Collaboration has reported the detections of 
$\gamma$-ray emission 
from several SNRs \cite{W51C,CasA,W44,IC443,W28,W49B}. 
Middle-aged SNRs interacting with molecular clouds (MCs) constitute 
the dominant class of  $\gamma$-ray-luminous SNRs. 
Here we summarize the results for such MC-interacting SNRs 
obtained with the LAT and discuss the mechanisms by which 
$\gamma$-ray emission can be enhanced in such systems.

\section{Fermi-LAT Detections of SNRs Interacting with Molecular Clouds}

The \emph{Fermi} LAT Collaboration has so far reported the discoveries 
of spatially \emph{extended} $\gamma$-ray sources coincident 
with four remnants: 
 W51C \cite{W51C}, W44 \cite{W44}, 
 IC~443 \cite{IC443}, and W28 \cite{W28}. 
They are middle-aged ($\sim 10^4$ yr) 
remnants that appear to be interacting with molecular clouds. 
Figure~1 shows a collection of the 2--10 GeV $\gamma$-ray count maps 
obtained with 2.5 yr data of the LAT. 
For better spatial resolution,
only  front-converted events are used. 
Even without subtracting the intense Galactic diffuse emission, 
significant $\gamma$-ray emission clearly stands out in excess of 
the background diffuse emission. 
In each panel, superimposed is a  
 radio continuum map synthesized with the Very Large Array (VLA) 
 at a frequency of: 
(a) 1.4 GHz \cite{KM97a}; (b) 1.4 GHz (The FIRST Survey); 
(c) 330 MHz \cite{Hewitt06}; and (d) 330 MHz.
The spatial extent of the $\gamma$-ray emission estimated through 
the maximum likelihood technique 
was generally found to be consistent with the size of the radio remnant. 
Given the size of a point-spread function of the LAT in this energy band, 
dedicated  investigations are 
 necessary to establish possible sub-structures within these remnants, 
 which will be reported elsewhere. 

\begin{figure}
\begin{center}
\includegraphics[width=0.8\textwidth]{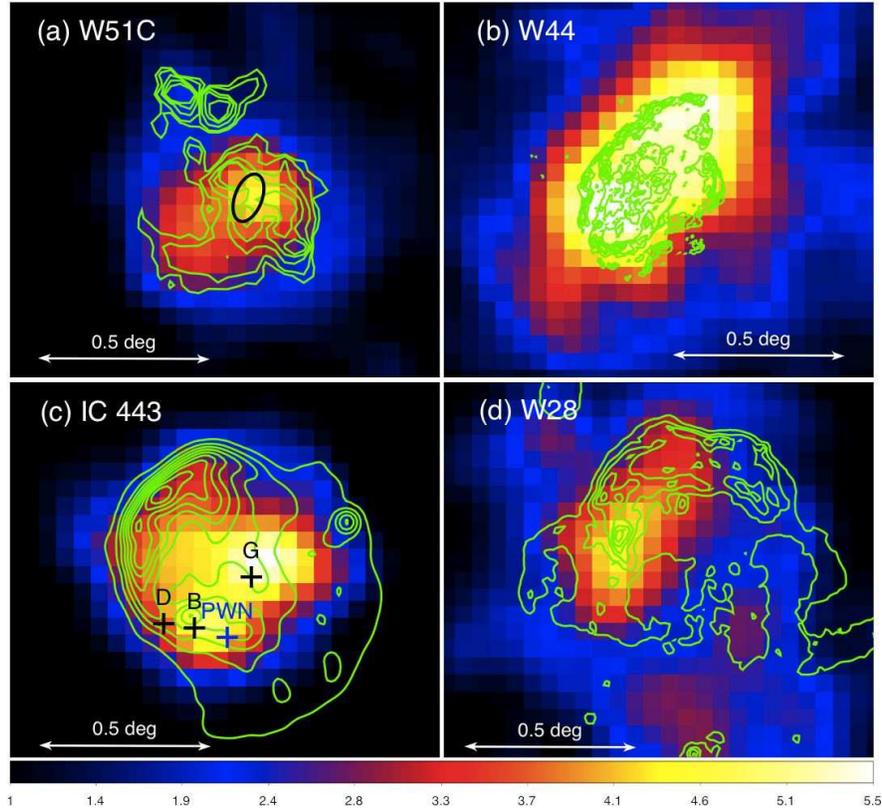}
\caption{
\emph{Fermi} LAT count maps in 2--10 GeV of 
 the MC-interacting SNRs with extended 
 $\gamma$-ray emission: (a) W51C; (b) W44; (c) IC443; and (d) W28. 
The LAT count maps (in the celestial coordinates) 
constructed from front-converted events 
 in the four panels are smoothed 
by a Gaussian kernel of 0.15 deg.
The intensity 
is  in units of counts per pixel with a pixel size of $0.05\times 0.05$~deg$^2$, 
and shown in a common range of 1.0--5.5 counts per pixel. 
Superposed are the contours from the VLA radio maps (see the text). 
A  black ellipse in panel (a) 
represents the location of shocked CO clumps \cite{KM97b}. 
The black crosses in panel (c) are the locations of  shocked molecular clumps 
from which OH maser is detected \cite{Hewitt06}. The position of 
a pulsar wind nebula is also marked in panel (c). 
}
\end{center}
\label{cmap}
\end{figure}

\begin{figure}
\begin{center}
\includegraphics[width=0.7\textwidth]{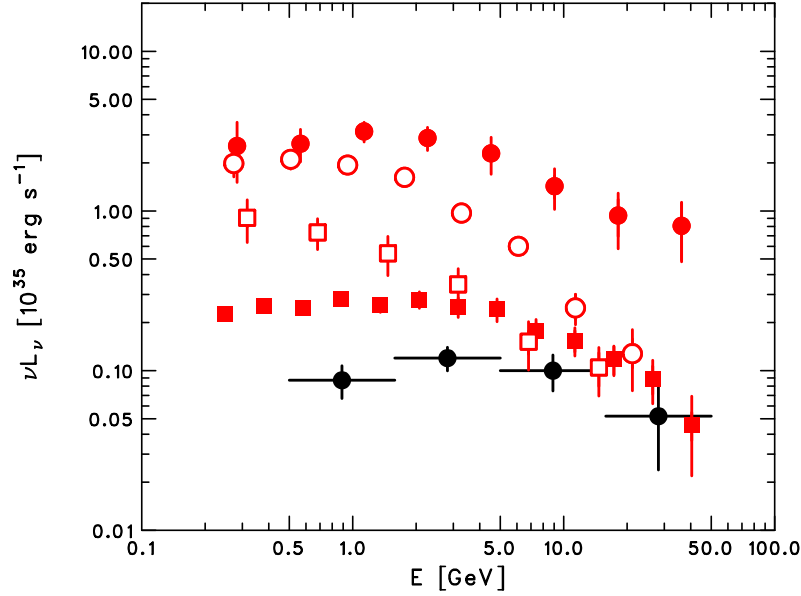}
\caption{Gamma-ray spectra of shell-type SNRs measured 
with the \emph{Fermi} LAT: W51C (red filled circles \cite{W51C}); 
W44 (red open circles \cite{W44}); 
IC~443 (red filled rectangles \cite{IC443}); 
W28 (red open rectangles \cite{W28}); 
Cassiopeia~A (black filled circles \cite{CasA}). 
}
\end{center}
\label{SED}
\end{figure}

The radio and $\gamma$-ray properties of the four SNRs are 
similar in many aspects. 
The synchrotron radio emission has a large flux of 160--310 Jy at 1 GHz 
with flat spectral index of $\alpha \simeq 0.3\mbox{--}0.4$ \cite{Green}. 
The GeV $\gamma$-ray spectrum commonly exhibits 
a spectral break at around 1--10 GeV, and the luminosity ranges 
$L_\gamma = 10^{35\mbox{--}36}\, \rm erg\ s^{-1}$. 
In Fig.~\ref{SED}, the $\gamma$-ray spectra of the MC-interacting SNRs 
 measured with the LAT are shown in the so-called $\nu L_\nu$ form 
along with the spectrum of  Cas~A. 
The distances are taken from the nominal values described in 
Refs.\ \cite{W51C,CasA,W44,IC443,W28}.
Note that the $\gamma$-ray luminosities of W51C and W44 are more than an order of magnitude larger than the $\gamma$-ray luminosity of Cas~A.

\begin{figure}
\includegraphics[width=0.5\textwidth]{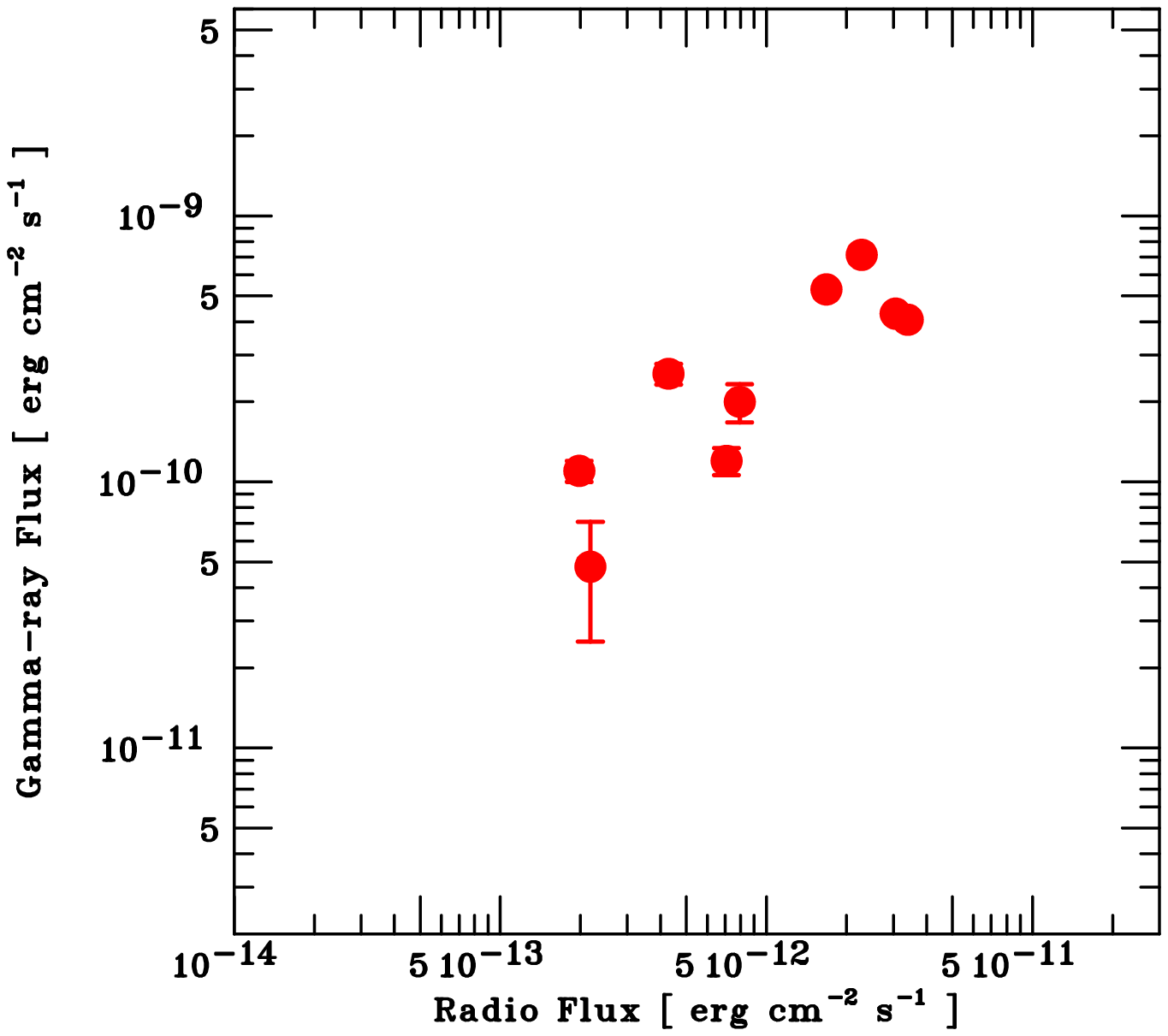}
\includegraphics[width=0.5\textwidth]{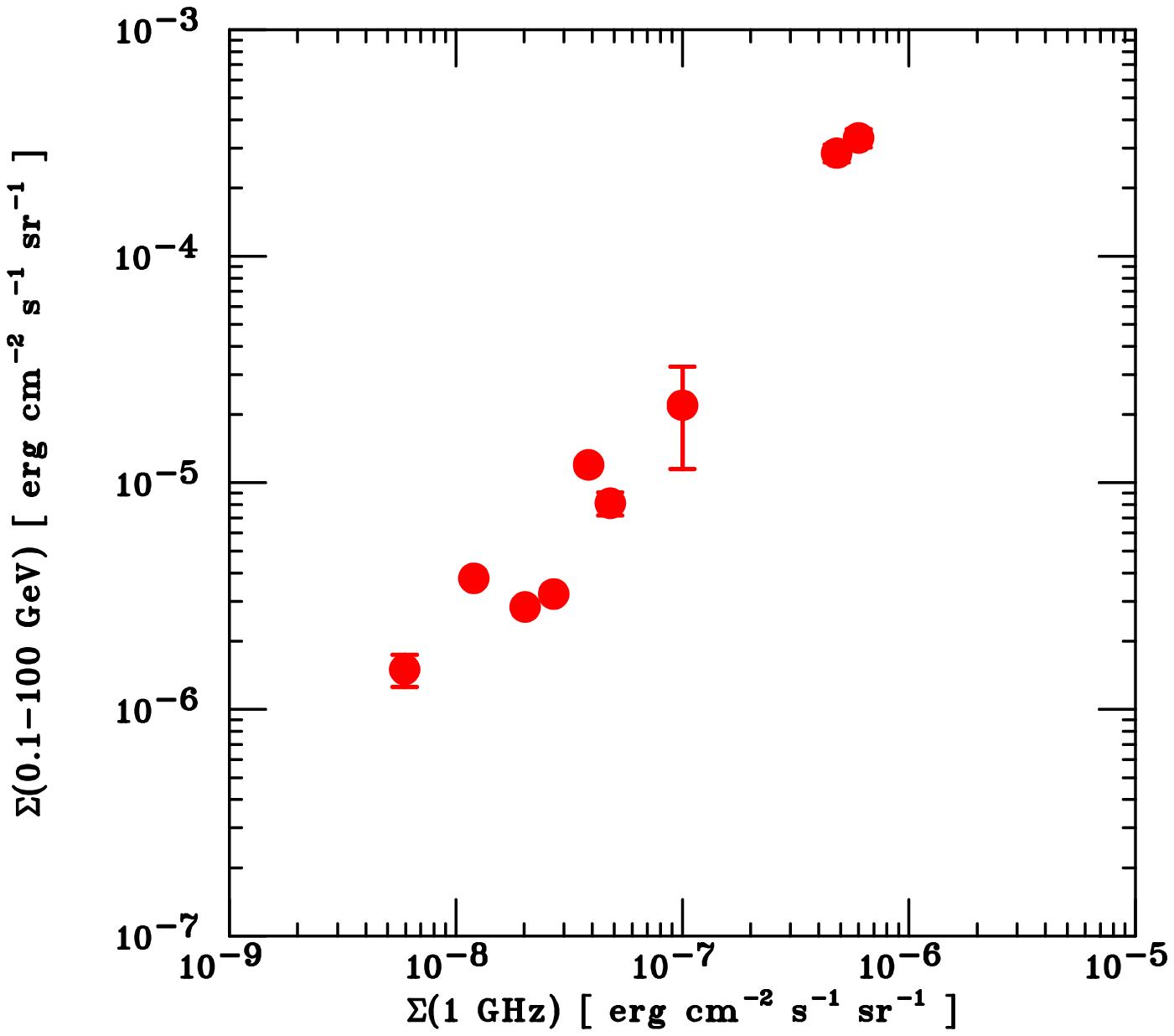}
\caption{
(Left) 
Radio flux (synchrotron) vs  GeV $\gamma$-ray flux for MC-interacting SNRs. 
The $\gamma$-ray energy flux integrated over 0.1--100 GeV and 
 the radio flux, $\nu f_\nu $ at 1 GHz, are shown. 
(Right) Mean surface brightness of the synchrotron radio emission 
and GeV $\gamma$-ray emission. The flux-flux plot is converted into 
this form using the solid angles of the radio remnants. }
\label{RadioGamma}
\end{figure}

The LAT-detected SNRs are generally radio-bright objects, which 
suggests some physical link between the synchrotron radiation and 
the GeV $\gamma$-rays. 
Indeed, the median value of the radio surface brightness\footnote{Usually, the radio surface brightness 
is represented as $\Sigma_\nu \equiv f_\nu /\Omega$ in units of 
$\rm W\ m^{-2}\ Hz^{-1}\ sr^{-1}$. 
Here we adopt $\Sigma \equiv \nu\Sigma_\nu = \nu f_\nu /\Omega$ in units of 
$\rm erg\ cm^{-2}\ s^{-1}\ sr^{-1}$.} at 1 GHz
of the Galactic SNRs
is $\Sigma \simeq 2 \times 10^{-9}\ \rm erg\ cm^{-2}\ s^{-1}\ sr^{-1}$ \cite{Green},
which is an order of magnitude smaller than that of the LAT-detected SNRs. 
Moreover, 
there found a possible correlation between the radio and $\gamma$-ray fluxes
from MC-SNRs. 
Figure~\ref{RadioGamma} (left) shows a plot of the radio flux vs $\gamma$-ray flux for the current list of MC-SNRs that are detected by 
the LAT \cite{W51C,CasA,W44,IC443,W28,W49B,Castro}. 
The radio and $\gamma$-ray energy fluxes represent the total fluxes 
integrated over a remnant. 
In Fig.~\ref{RadioGamma} (right), 
a comparison between the radio and $\gamma$-ray intensities 
is made in terms of 
mean surface brightness, where the total flux is divided by 
the solid angle subtended by each remnant. 
The mean surface brightness, which is distance-independent, 
has a large dynamic range over more than two orders of magnitude.


\section{Discussion}

Since the interaction with a molecular cloud clearly plays a key role in enhancing 
the $\gamma$-ray emission,  bremsstrahlung by relativistic electrons 
or $\pi^0$-decay $\gamma$-rays produced by high-energy protons are 
the two plausible dominant channels of the $\gamma$-radiation. 
The observed large luminosity of the GeV $\gamma$-ray emission precludes 
the inverse-Compton scattering off the CMB and interstellar radiation fields 
as the main emission mechanism. For instance, 
the luminosity of SNR W51C requires an unacceptably large amount of 
the total CR electron energy of 
$W_e \simeq 1\times 10^{51}\ \rm ergs$ \cite{W51C}.
Generally the high-energy protons are thought to be the 
predominant component of accelerated particles in SNRs so that 
the $\pi^0$-decay $\gamma$-ray emission is stronger than 
the electron bremsstrahlung (see however Ref.\ \cite{Bykov}). 

There are two different types of scenarios to explain the GeV $\gamma$-ray 
emission arising from the MC-SNR systems.
The {\it Runaway CR} model 
considers  $\gamma$-ray emission from 
molecular clouds illuminated by runaway CRs 
that have escaped from their accelerators \cite{AA96,Gabici09,Ohira11}. 
CR particles released by a remnant  diffuse into interstellar space 
with an energy-dependent diffusion coefficient, and collide with 
massive clouds producing $\gamma$-ray photons via 
$\pi^0$-decay (by CR hadrons) and bremsstrahlung (by CR electrons) channels. 
The $\gamma$-ray wide-band spectra from the GeV to TeV energies 
can  markedly  vary 
from one object to another depending on the relative locations of the accelerator 
and the target cloud as well as  the time history of the accelerator. 
 For example, 
a hard GeV $\gamma$-ray source \cite{W28} 
which is also bright in TeV \cite{HESSW28} 
outside the southern boundary of SNR W28 may be explained by 
runaway CRs. 
Future analysis of spatial distributions of GeV $\gamma$-rays 
combined with CO observations 
would make it possible to discern the $\gamma$-ray sources that are attributable 
to runaway CRs. 
Then one can learn about how CRs are released into interstellar space and 
how they propagate in the vicinity of the SNR where 
self-generated Alfv\'en waves may change the local diffusion coefficient
\cite{Fujita10}.

Another scenario, the {\it Crushed Cloud} model \cite{Uchi10} 
invokes a  ``shocked" molecular cloud into which a radiative shock 
(typically $v_{\rm s} \sim 100\ \rm km\ s^{-1}$) 
is driven by the high pressure of hot plasmas in the blastwave region. 
The cosmic-ray particles accelerated at a cloud shock are 
adiabatically compressed behind the shock front, resulting in enhanced 
synchrotron and $\pi^0$-decay $\gamma$-ray emissions \cite{BC82}.
It was shown by Ref.~\cite{Uchi10} that reacceleration of 
pre-existing CR electrons and protons is capable of explaining both radio and GeV $\gamma$-ray spectra  using reasonable sets of 
parameters. 
The $\gamma$-ray luminosity of  $L_\gamma \sim 10^{35}\, \rm erg\ s^{-1}$ 
in 1--10 GeV agrees well with the theoretical expectation. 
In this scenario, the radio and $\gamma$-ray emissions from 
the MC-SNRs are intimately connected, and therefore the 
possible radio-gamma correlation can be readily explained. 

The Runaway CR and Crushed Cloud models are not mutually exclusive. 
The former presumes unshocked molecular clouds as the 
$\gamma$-ray production sites while the latter considers shocked 
clouds. 
Both unshocked and shocked clouds could be the important sites 
of the $\gamma$-ray production. 
Also, in addition to the pre-existing GCRs in a molecular cloud, 
runaway CRs released from a shock may make a large contribution to the 
cosmic-ray density in the molecular cloud particularly at TeV energies. 
When such a cloud is struck and compressed by the blastwave, 
the $\gamma$-ray flux at high energies should be enhanced with 
respect to what is presented in the model of Ref.\ \cite{Uchi10}.

\acknowledgments
The $Fermi$ LAT Collaboration acknowledges support from a number of agencies and institutes for both development and the operation of the LAT as well as scientific data analysis. These include NASA and DOE in the United States, CEA/Irfu and IN2P3/CNRS in France, ASI and INFN in Italy, MEXT, KEK, and JAXA in Japan, and the K.~A.~Wallenberg Foundation, the Swedish Research Council and the National Space Board in Sweden. Additional support from INAF in Italy and CNES in France for science analysis during the operations phase is also gratefully acknowledged.


\begin{thebibliography}{99}

\bibitem{BE87}
Blandford, R., \& Eichler, D.,
\emph{Particle acceleration at astrophysical shocks: A theory of cosmic ray origin},
Physics Reports, 154, 1-75, 1987.

\bibitem{Reynolds08} 
Reynolds, S.P., 
\emph{Supernova Remnants at High Energy}, 
ARAA, 46, 89-126, 2008. 

\bibitem{Uchiyama07} 
Uchiyama, Y., Aharonian, F.A., Tanaka, T., et al., 
\emph{Extremely fast acceleration of cosmic rays in a supernova remnant}, 
Nature, 449, 576-578, 2007. 

\bibitem{HESS04}
Aharonian, F.A., Akhperjanian, A.G., Aye, K.M., et al., 
\emph{High-energy particle acceleration in the shell of a supernova remnant},
Nature, 432, 75-77, 2004.

\bibitem{Esposito}
Esposito, J.A., Hunter, S.D., Kanbach, G., Sreekumar, P.,
\emph{EGRET Observations of Radio-Bright Supernova Remnants},
ApJ, 461, 820-827, 1996.

\bibitem{Torres}
Torres, D.F., Romero, G.E., Dame, T.M., et al.,
\emph{Supernova remnants and gamma-ray sources}, 
Phys.\ Rep., 382, 303-380, 2003.

\bibitem{Fermi-Pulsars}
Abdo, A.A., Ackermann, M., Ajello, M., et al., 
\emph{The First Fermi Large Area Telescope Catalog of Gamma-ray Pulsars}, 
ApJS, 187, 460-494, 2010.

\bibitem{LAT}
Atwood, W.B., Abdo, A.A., Ackermann, M., et al.,
\emph{The Large Area Telescope on the Fermi Gamma-Ray Space Telescope Mission},
ApJ, 697, 1071-1102, 2009.

\bibitem{W51C}
Abdo, A.A., Ackermann, M., Ajello, M., et al., 
\emph{Fermi LAT Discovery of Extended Gamma-Ray Emission in the Direction of Supernova Remnant W51C},
ApJL, 706, 1-6, 2009.

\bibitem{CasA}
Abdo, A.A., Ackermann, M., Ajello, M., et al., 
\emph{Fermi-LAT Discovery of GeV Gamma-Ray Emission from the Young Supernova Remnant Cassiopeia A},
ApJL, 710, 92-97, 2010.

\bibitem{W44}
Abdo, A.A., Ackermann, M., Ajello, M., et al., 
\emph{Gamma-Ray Emission from the Shell of Supernova Remnant W44 Revealed by the Fermi LAT}, 
Science, 327, 1103-1106, 2010.

\bibitem{IC443}
Abdo, A.A., Ackermann, M., Ajello, M., et al., 
\emph{Observation of Supernova Remnant IC 443 with the Fermi Large Area Telescope}, 
ApJ, 712, 459-468, 2010.

\bibitem{W28}
Abdo, A.A., Ackermann, M., Ajello, M., et al., 
\emph{Fermi Large Area Telescope Observations of the Supernova Remnant W28 (G6.4-0.1)},
ApJ, 718, 348-356, 2010.

\bibitem{W49B}
Abdo, A.A., Ackermann, M., Ajello, M., et al., 
\emph{Fermi-LAT Study of Gamma-ray Emission in the Direction of Supernova Remnant W49B}, 
ApJ, 722, 1303-1311, 2010.

\bibitem{KM97b} 
Koo, B.-C., \& Moon, D.-S. 
\emph{Interaction between the W51C Supernova Remnant and a Molecular Cloud. II. Discovery of Shocked CO and HCO+}, 
ApJ, 485, 263-269, 1997. 

\bibitem{Hewitt06}
Hewitt, J.W., Yusef-Zadeh, F., Wardle, M., et al., 
\emph{Green Bank Telescope Observations of IC 443: The Nature of OH (1720 MHz) Masers and OH Absorption}, 
ApJ, 652, 1288-1296, 2006. 

\bibitem{KM97a} 
Koo, B.-C., \& Moon, D.-S.,
\emph{Interaction between the W51C Supernova Remnant and a Molecular Cloud. I. HI 21 Centimeter Line Observations}, 
ApJ, 475, 194-210, 1997.

\bibitem{Green}
Green, D.A.,
\emph{A revised Galactic supernova remnant catalogue},
Bulletin of the Astronomical Society of India, 37, 45-61, 2009.

\bibitem{Castro}
Castro, D., \& Slane, P.,
\emph{Fermi Large Area Telescope Observations of Supernova Remnants Interacting with Molecular Clouds},
ApJ, 717, 372-378, 2010.

\bibitem{Bykov}
Bykov, A.M., Chevalier, R.A., Ellison, D.C., \& Uvarov, Y.A.,
\emph{Nonthermal Emission from a Supernova Remnant in a Molecular Cloud}, 
ApJ, 538, 203-216, 2000.

\bibitem{AA96}
Aharonian, F.A., \& Atoyan, A.M.,
\emph{On the emissivity of $\pi^0$-decay gamma radiation in the vicinity of accelerators of galactic cosmic rays},
A\&A, 309, 917-928, 1996.

\bibitem{Gabici09}
Gabici, S., Aharonian, F.A., \& Casanova, S.,
\emph{Broad-band non-thermal emission from molecular clouds illuminated by cosmic rays from nearby supernova remnants},
MNRAS, 396, 1629-1639, 2009.

\bibitem{Ohira11}
Ohira, Y., Murase, K., \& Yamazaki, R., 
\emph{Gamma-rays from molecular clouds illuminated by cosmic rays escaping from interacting supernova remnants},
MNRAS, 410, 1577-1582, 2011. 

\bibitem{HESSW28}
Aharonian, F., Akhperjanian, A.G., Bazer-Bachi, A.R., et al.,
\emph{Discovery of very high energy gamma-ray emission coincident with molecular clouds in the W 28 (G6.4-0.1) field}, 
A\&A, 481, 401-410, 2008.

\bibitem{Fujita10}
Fujita, Y., Ohira, Y., \& Takahara, F.,
\emph{Slow Diffusion of Cosmic Rays Around a Supernova Remnant}, 
ApJL, 712, 153-156, 2010.

\bibitem{Uchi10}
Uchiyama, Y., Blandford, R., Funk, S., Tajima, H., \& Tanaka, T.,
\emph{Gamma-ray Emission from Crushed Clouds in Supernova Remnants},
ApJL, 723, 122-126, 2010.

\bibitem{BC82}
Blandford, R., \& Cowie, L.,
\emph{Radio emission from supernova remnants in a cloudy interstellar medium},
ApJ, 260, 625-634, 1982.


\end{thebibliography}
\end{document}